 \documentclass[conference]{IEEEtran}

\usepackage{ifpdf}

\usepackage{cite}

  \usepackage{graphicx}
   \DeclareGraphicsExtensions{.pdf,.jpeg,.png}

\usepackage[cmex10]{amsmath}
\usepackage{array}

\usepackage{eqparbox}
\usepackage[caption=false]{subfig}

\usepackage{dblfloatfix}

\usepackage{url}

\usepackage[autostyle]{csquotes}
\usepackage{mdwlist}
\newcommand{\bit}{\begin{itemize}}
\newcommand{\eit}{\end{itemize}}
\usepackage{amsfonts}
\usepackage{amssymb}
\usepackage{verbatim}

\usepackage[table]{xcolor}
\usepackage{multirow}

\newcolumntype{a}{>{\columncolor{lightgray}}c}

\begin{document}
%
\title{Rapid Sequence Identification of Potential Pathogens Using Techniques from Sparse Linear Algebra}

\author{\IEEEauthorblockN{Stephanie Dodson, Darrell O. Ricke, Jeremy Kepner, Nelson Chiu, and Anna Shcherbina}
\IEEEauthorblockA{MIT Lincoln Laboratory, Lexington, MA, U.S.A\\
}}


%


\maketitle

\begin{abstract}
The decreasing costs and increasing speed and accuracy of DNA sample collection, preparation, and sequencing has rapidly produced an enormous volume of genetic data. However, fast and accurate analysis of the samples remains a bottleneck. Here we present D$^{4}$RAGenS, a genetic sequence identification algorithm that exhibits the Big Data handling and computational power of the Dynamic Distributed Dimensional Data Model (D4M).  The method leverages linear algebra and statistical properties to increase computational performance while retaining accuracy by subsampling the data.  Two run modes, Fast and Wise, yield speed and precision tradeoffs, with applications in biodefense and medical diagnostics. The D$^{4}$RAGenS analysis algorithm is tested over several datasets, including three utilized for the Defense Threat Reduction Agency (DTRA) metagenomic algorithm contest.  
\end{abstract}


%
\IEEEpeerreviewmaketitle
\let\thefootnote\relax\footnote{This work is sponsored by the Assistant Secretary of Defense for Research \& Engineering under Air Force Contract \#FA8721-05-C-0002. Opinions, interpretations, recommendations and conclusions are those of the authors and are not necessarily endorsed by the United States Government.}

\section{Introduction}

With advances in biological and chemical knowledge as well as increased world travel, bioterrorism and deadly disease outbreaks are real concerns. The mitigation of these threats lies in accurate source and organism identification, making rapid genetic sequencing and organism identification of particular interest in homeland defense.  

Bioterror attacks made headlines in 2001 with the mailing of anthrax containing letters to notable media and political figures that resulted in 17 cases and 5 deaths. The timing of the first letter, shortly after the September 11$^{th}$ terror attacks, increased public anxiety of the possibility of a foreign attack \cite{NY_times_antrax_public_concern}. A nine-year F.B.I. investigation ensued that dismissed the idea of a foreign 
assailant when analysis linked the anthrax strain and concentration to those previously developed by the United States biowarfare defense program \cite{anthrax_linked_to_US}. Finally, a U.S. government scientist was indicted for the crimes \cite{npr_anthrax_timeline}. 

An example of widespread food borne illness lies in the 2011 virulent Germany \emph{E. coli} outbreak.  A slightly pathogenic organism, \emph{E. coli} typically has minor gastrointestinal effects and little chance of lasting damage or death. The Germany outbreak however was much more serious and over the course of almost three months, the bacteria swept through Germany affecting over 4,000 individuals and resulting in 53 fatalities \cite{EHEC_report}. Genetic sequencing did play a role in the discovery of a bacteriophage insertion carrying the Shiga toxin in the \emph{E. coli} strain O104:H4 that accounted for the severity \cite{EHEC_report}. 

Despite the sequencing, the source was not correctly identified as sprouts until 41 days after the first hospitalized case, and in the mean time the media and critics ran wild with the possibility that the bacteria was a genetically engineered bioattack \cite{ecoli_bioattack}.  Additionally, the misidentification of the source as Spanish cucumbers caused European Union farmers weekly losses up to \$611 million \cite{cucumber_losses}.

In 2001 and 2011 genetic sequencing machines existed, but the lack of widely available fast analysis techniques inhibited the identification of the source of the \emph{E. coli} and anthrax.  Since the anthrax attacks the United States has developed federal laboratories to counter biological terrorism, but fast detection and characterization methods are still inadequate \cite{detection_methods_inadequate}. 

Sequencing machines continue to be upgraded. At a fraction of the previous cost, new Next Generation Sequencers (NGS) have the capability of sequencing an entire human genome in mere days, with smaller genomes in minutes to hours. These sequencers combined with fast analytic methods will allow for fast detection methods with endless application in homeland protection as well as infectious disease, inherited disease, cancer genomics, and individualized medicine. 

In general, metagenomics algorithms seek to tackle four analysis problems: (1) organism identification, (2) correct classification of sequences, (3) gene identification, and (4) variant identification. With the goal of finding an algorithm that would allow DNA sequencing to become a rapid diagnostic medical tool, the Defense Treat Reduction Agency (DTRA) sponsored a challenge to find a fast and accurate method\cite{DTRA_contest}.  Several metagenomic algorithms are currently available, some as a result of the DTRA challenge. Each algorithm uses a wide range of techniques, with trade-offs between time and accuracy. 

Many algorithms utilize existing tools, such as the \enquote{gold-standard} BLAST (Basic Local Alignment Search Tool), and modify the results to achieve better identification levels \cite{metaphyler}\cite{metaphlan}\cite{megan}. These algorithms, including MetaPhlyer \cite{metaphyler} and MetaPhlAn \cite{metaphlan}, can achieve faster speeds by  running BLAST searches on reduced datasets of marker genes. Other algorithms test sequence similarity through the use of computationally expensive hidden-Markov-models (HMMs) and creation of phylogenetic trees \cite{sub_hmm}\cite{mltree}.  

To avoid cumbersome external programs, HMMs, and phylogenetic trees, \cite{k_mer} instead segments the sequences into short k-mers, and performs sequence comparisons by matching the k-mers in a sorted list. The methods MetaCV \cite{metacv}, LMAT \cite{lmat}, and Kraken \cite{kraken} employ a similar k-mer segmenting strategy in combination with phylogenetic trees to discover the least common ancestors. 

In this paper we present D$^{4}$RAGenS, a simplified k-mer comparison style algorithm that leverages the sparse linear algebra and computational power of the Dynamic Distributed Dimensional Data Model (D4M).  First defined in \cite{D4M_gen_seq_match}, D$^{4}$RAGenS (D4M Rapid Analysis of Genetic Sequences) has been modified to include the data subsampling performance techniques described in \cite{Taming_Big_Data}. The key idea is the segmentation of genetic sequences into short k-mers for easy and efficient string comparison with a basic matrix multiplication in D4M. 

Furthermore, with two run modes, Fast and Wise, D$^{4}$RAGenS undertakes the problems of organism identification and read classification resulting in a simple, but rigorous detection method that scales with increasing data volume.  Results from several \emph{in silico} test datasets are compared with current algorithms.


\section{Methods}

\subsection{D4M}

Developed at MIT Lincoln Laboratory, D4M is an environment for Matlab that blends techniques from sparse linear algebra, graph theory, and abstract algebra to create triple-store format associative arrays. In the triple-store format, arrays can have strings and/or numerics as the row, column, and value keys, allowing for easy data querying. The structure of arrays allows for easy parallelization and increased computation capacity with minimal extra code. Additionally, all standard mathematical operations of multiple associative arrays are composable, and result in an associative array \cite{D4M}. 

The architecture of D4M provides the tools necessary to create a rapid sequence alignment algorithm that takes advantage of simple mathematical properties. 

\begin{figure}[!t]
\centering
\includegraphics[width=2.8in]{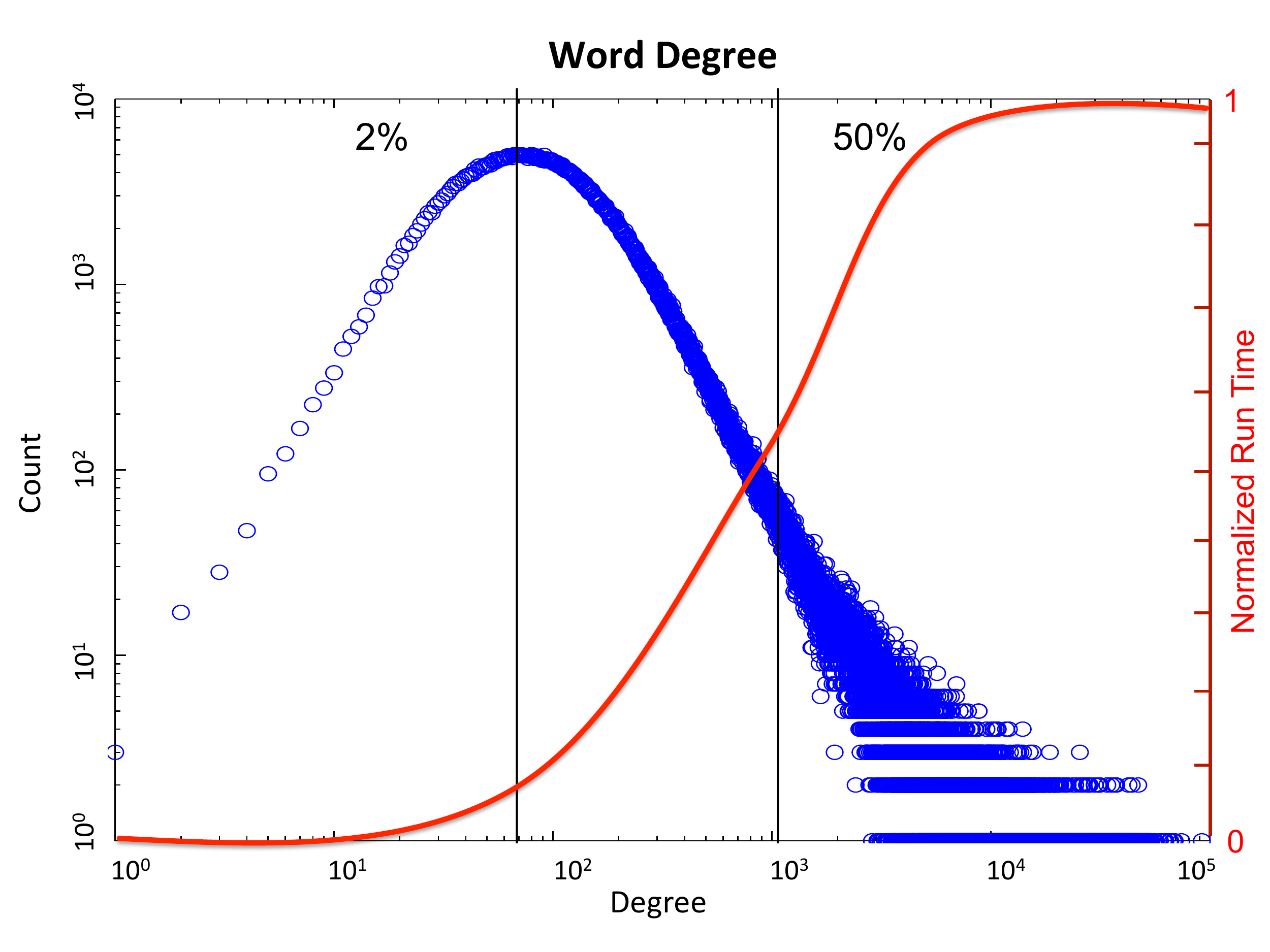}
\caption{Degree distribution of the unique k-mers in the unique bacterial RNA sequences. Vertical lines indicate the percent of data distribution. The data is dominated by few common words, or \enquote{super-nodes.}  Increasing the percentage of data used drastically increases the run time as shown by the red curve.}
\label{10mer_distribution}
\end{figure}

\subsection{Sequence comparison}

During processing, NGS systems segment input DNA sequences into a relatively short length (typically 150-450 base pairs (bp) long). The short segments, called reads, are amplified to increase concentration, and then the DNA sequence is determined.  Sequences are then characterized by comparisons to known DNA, RNA, and protein sequences. DNA and RNA sequence comparisons are the focus of this paper. 

Comparing just two sequences as strings is typically an order N$^{2}$ computational problem. An alternative to string comparison is converting sequences into sparse vector representations and using the mathematical dot product operation as a measure of identity. Sequences are split into non-redundant overlapping 10 base pair (bp) k-mers, or words and used as vector indices.  A vector dot produce yields the number of exact word matches between the two sequences. Furthermore, parallel comparisons between K unknown sequences of length N to M reference sequences (order (K x M) x N$^{2}$) is computed with only a sparse matrix multiplication. 

Word length is chosen to optimize the algorithm sensitivity and array sparseness. The four DNA bases and 10 bp per word leads to a domain of 4$^{10}$ (1,048,576) unique words. Long DNA sequences are divided into segments of 1,000 bp with a 100 bp overlap prior to word formation. This maximum segment length and domain size leads to a 1 in 1,000 chance of random word matches between unrelated sequences.


\subsection{Subsampling of k-mers}

\begin{figure}[!t]
\centering
\includegraphics[width=3in]{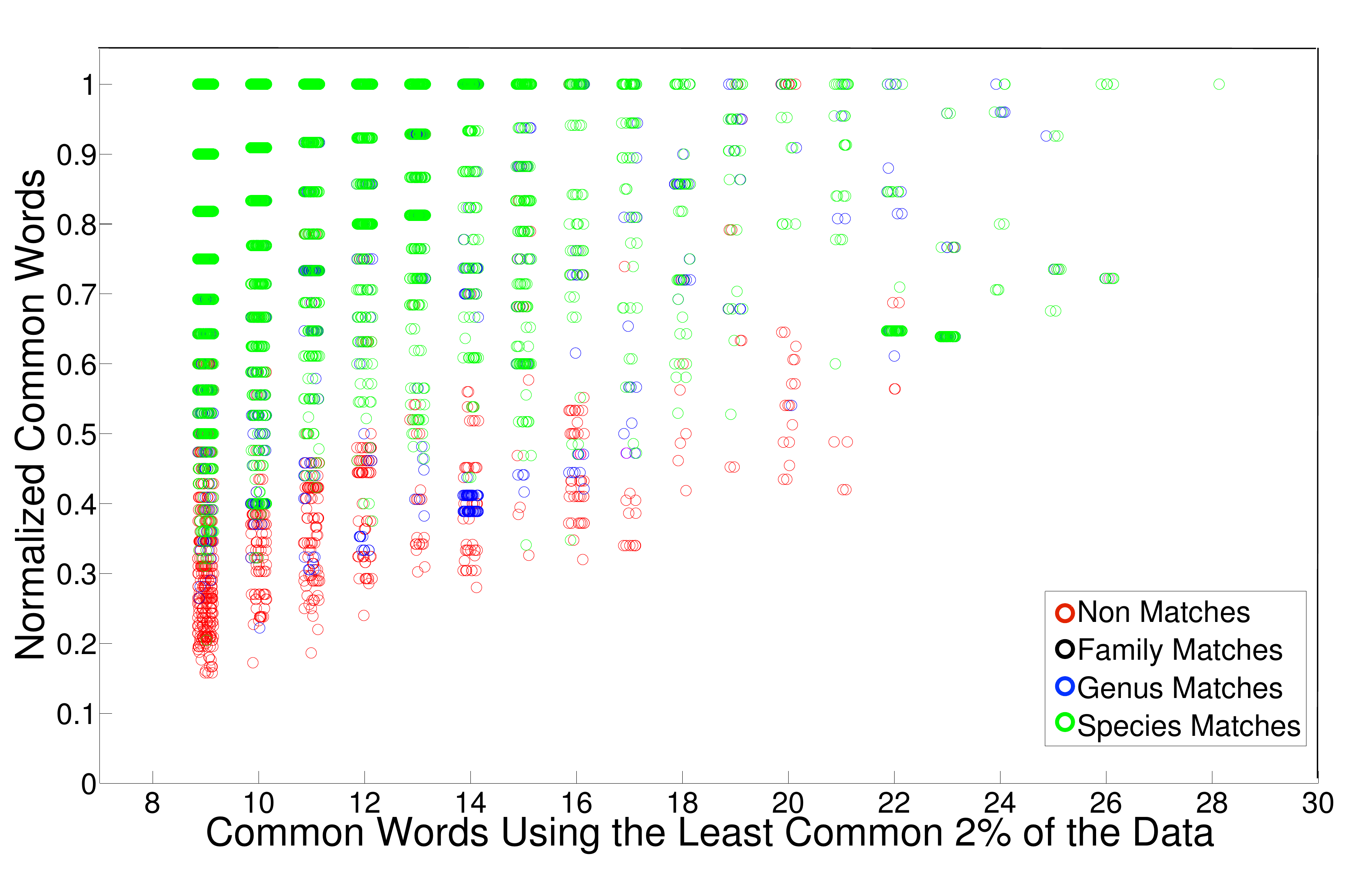}
\caption{Normalizing the hit counts by the subsampled sequence length acts as a filter for signal and noise. Matches resulting in true organism identification (green points) have high normalization values, while false positives (red points) are clustered around low normalized values. }
\label{normalization}
\end{figure}

The D4M graph-linear algebra duality allows for the creation of a bipartite graph between unique words and sequence identifiers. An edge exists between the vertex sets if that word is contained in the sequence, and the word degree identifies the frequency of use.  In random sequences, all words occur with the same frequency. DNA however, is not random and common words represent highly conserved sequences that are present in many organisms (e.g. ribosomal RNA genes). The nonrandom and repetitive nature of DNA is highlighted in the degree distribution of the words (Figure \ref{10mer_distribution}), which shows the data is dominated by few words of high frequency.

Frequent word use hampers the identification performance with extraneous data, numerous false positives, and long computation times. Removal of these supernodes results in the dramatic reduction of comparisons between the reference dataset and unknown sample, resulting in decreased run time. The red curve in Figure \ref{10mer_distribution} shows the algorithm run time as a function of the percent of data. Run time increases quickly with data usage.  Results of \emph{in silico} generated datasets show the organism identification power is retained by measuring sequence identity (performing the sparse matrix multiplication) with the least common 2\% of the data \cite{ICASSP}.


\subsection{Sequence normalization}

Sequence similarity computed with the least common words discovers the truth organisms (based on the addition of FastqSim \cite{FastqSim} \emph{in silico} generated sequences to the datasets), but the results are plagued by false positive identifications caused by the reduced sample space. Normalizing the number of common words by the length of the subsampled sequence acts a second noise filter, as shown in Figure \ref{normalization}.


\section{Algorithms}

\subsection{Fast}

The Fast algorithm combines the data subsampling with sequence length normalization to rapidly identify the organisms present in the sample. After subsampling, the unknown sequences and reference database are compared with a sparse matrix multiplication. 

The words in common are first thresholded to eliminate chance k-mer matches, before selecting matches with normalization above 0.9. In the case of multiple alignments, the sample is classified to the least common taxonomic similarity between all strong reference matches. 

The two filters work together to remove extraneous false matches. Weaker true matches are also removed, but the strongest matches remain and allow organism detection even for low read numbers. Computational performance is increased by parallelizing the sequence comparisons and length normalizations.

\begin{figure*}[!t]
\centering
\includegraphics[width=5.36in]{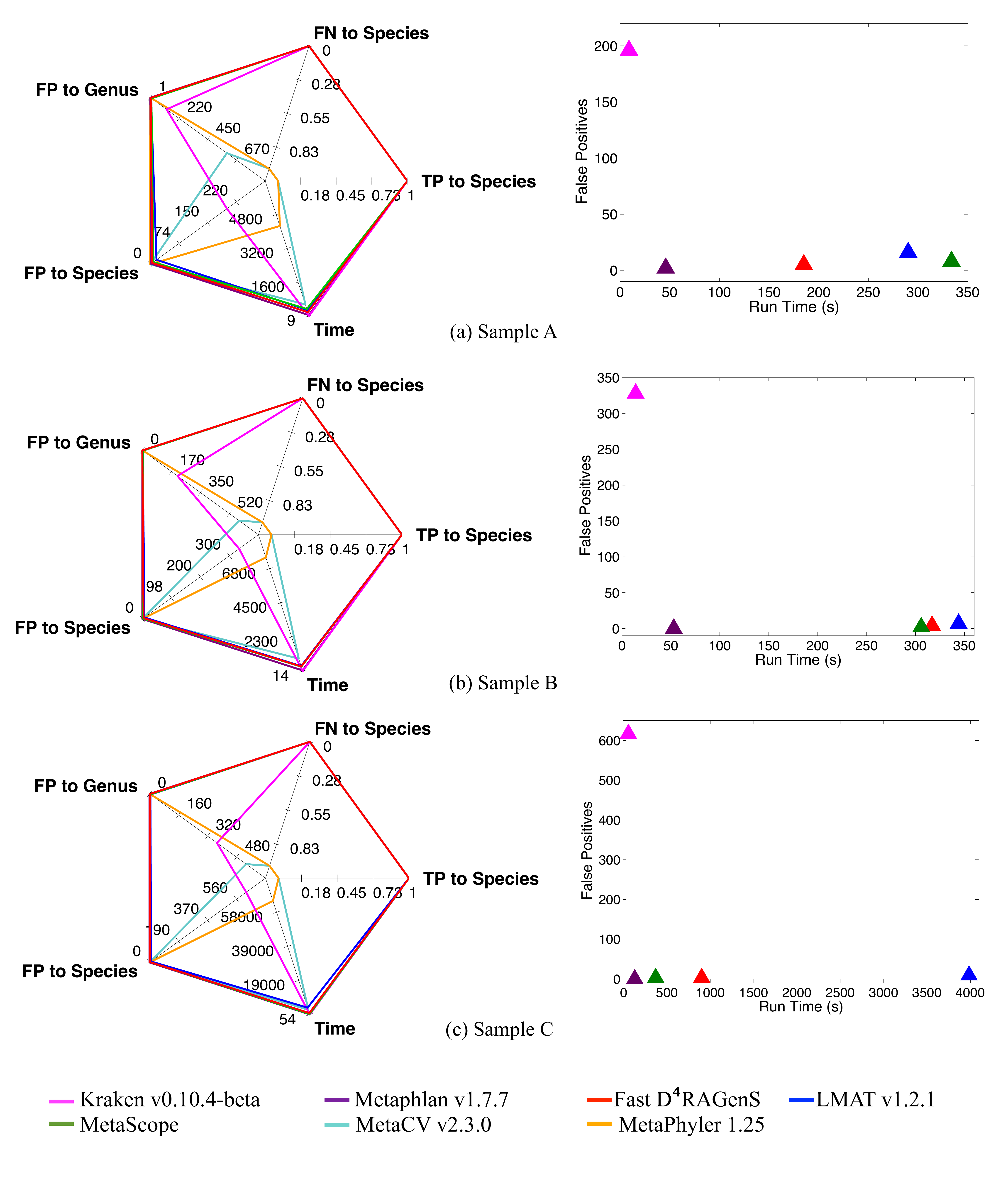}
\caption{Organism identification results for samples A, B, and C. The left plots display the identification power of the algorithms. Each axis represents a testing parameter used in the comparison. The charts are designed with each axis radiating outwards from poor to good results. For example, high numbers of true positives are desired, thus the true positive axes begin with 0 at the center and increase along the axis length. False positives are the opposite, and begin with high numbers at the center and decrease outward to 0. Thus, strongest candidates lie on the perimeter of the chart. The number of false positive species versus run times for the fastest five algorithms are highlighted in the figures on the right. }
\label{algorithm_comparisons_T3_T4_T6}
\end{figure*}


\subsection{Wise}

The Wise algorithm expands upon Fast mode to identify the organism and correctly classify reads. Fast mode is used as a first pass filter to identify target organisms. Then, through a second matrix multiplication, the sample is compared to the identified target organisms, this time using the full distribution of words to maximize the matching power. Common words are again thresholded to reduce noise.  

Preselecting target organisms condenses the reference set, allowing for a quick multiplication with the full word domain and minimizing noise caused by chance matches.  The final common word threshold helps to reduce noise caused by family and genus level matches detected by Fast mode. Again, sequence comparison multiplications are performed in parallel to increase the comparison rate. 

For both Fast and Wise comparisons, the reference databases are built from data in the GenBank database. The RNA data is selected to provide a reduced dataset by selecting only the longest gene sequence for each organism by genus species. The selection eliminates redundancy and optimizes algorithm performance.


\section{Results}

The Fast and Wise modes were tested on several \emph{in silico} datasets and results compared with six competing algorithms. Organism identification and read mapping was tested using three \emph{in silico} samples of human data spiked with bacterial data  (samples A, B, and C) and one \emph{in silico} viral sample. D$^{4}$RAGenS was run using 60 cores on a Linux cluster with D4M. The other metagenomic algorithms were all run using 60 cores on an HP SMP server.

Sample A was processed on an Ion Torrent sequencer, and contains 379,027 total unknown sequences (reads) with an average length of 160 bp. Sample B was sequenced on the Roche platform and has a longer average read length of 363 bp and a total of 399,670 reads. Sample C is Illumina data, with 6,038,556 reads of length 150 bp.  The data size is predominately influenced by the number of reads, making sample C the largest. In addition to the human background, the three bacterial samples contain \emph{in silico} spiked bacterial organisms. 

These samples were used in the DTRA challenge, and the spiked organisms all have pathogenic properties that can be exploited by bioterrorists. \emph{Klebsiella pneumonia} in sample A has multi-drug resistance and is common within health care facilities \cite{Klebsiella}. The bacterial organisms in sample B and C have been identified as possible biological weapons that can threaten the food supply \cite{brucella} \cite{bacillus_anthracis}. Brucellosis, a disease caused by species in the \emph{Brucella} genus, including the \emph{Brucella abortus} present in sample B, is easily transmissible between humans and livestock, and causes undulant fever in humans \cite{brucella}. Sample C contains \emph{Bacillus anthracis}, the organism responsibe for the potentially fatal anthrax infection \cite{bacillus_anthracis}. 

The \emph{in silico} viral sample was created with FastqSim \cite{FastqSim} to simulate Illumina Sequencing data. The test set contains 22 known viruses at various concentrations, with average read length of 160 bp. Again, all viruses present commonly infect humans, some with fatal consequences, such as the \emph{Sudan ebolavirus}.


\subsection{Bacterial Organism Identification}

In organism identification, the goal is to rapidly distinguish the bacterial organisms from the human background and maintain a low rate of false positives. The algorithms were scored using five criteria: number of true positives to species, false positives to species and genus, false negatives, as well as total run time. Results of the seven algorithms on the bacterial samples A, B, and C are shown in the radar plots in Figure \ref{algorithm_comparisons_T3_T4_T6}. The arrangement is such to showcase the stronger algorithms along the perimeter of the chart. 

Five of the seven algorithms correctly identified the spiked bacterial organisms in the three samples and received highest marks for the true positive and false negative to species. The other two, MetaCV v2.3.0 and MetaPhlyer 1.25, are designed to identify only to the genus level and thus receive zero false positive species by default.  The usefulness of the top five algorithms is shown in the scores for the remaining categories: false positives to genus and species as well as run time. 

The Kraken v0.10.4-beta algorithm consistently has the fastest run time and identified the spiked organism, but reports an excessive amount of false positives. An additional 200-600 organisms were identified, yielding a non-specific answer. 

The run times and false positives to species of the top five algorithms are highlighted in the plots on the right of Figure \ref{algorithm_comparisons_T3_T4_T6}.  The four methods Metaphlan v1.7.7, Fast D$^{4}$RAGenS, MetaScope \cite{DTRA_contest}, and LMAT v1.2.1 have reasonable run times and false positives.  The Fast D$^{4}$RAGenS algorithm consistently places in the top three for both categories (excluding Kraken v0.10.4-beta due to high false positives).

\begin{table}[!t]
\caption{Number of reads classified as each truth organism by the six detection algorithms that report read mapping. The last row gives the results of running Wise mode on a gdna reference database for samples A and B. }
\label{table:bacteria_read_mapping}
\centering
\begin{tabular}{c|cc|cc|cc}
\hline
\textbf{Algorithm}&\multicolumn{2}{c|}{\textbf{SampleA}} & \multicolumn{2}{c}{\textbf{Sample B}} &\multicolumn{2}{|c}{\textbf{Sample C}} \\
\hline
&Species&Genus&Species&Genus&Species&Genus\\[1ex]
\hline
Truth & 58,707 &58,707 & 49,948 & 49,948 & 39,996 & 39,996\\ [1ex]
\hline
Wise  & \multirow{2}{*}{37,486} & \multirow{2}{*}{37,486} & \multirow{2}{*}{961} & \multirow{2}{*}{24,558} & \multirow{2}{*}{4,107} & \multirow{2}{*}{21,261}\\
D$^{4}$RAGenS &  & & & & &\\[1ex]
MetaScope& 53,999 & 54,109 & 48,675 & 49,517 & 39,613 & 39,867 \\[1ex]
Kraken & \multirow{2}{*}{49,158} & \multirow{2}{*}{51,995} & \multirow{2}{*}{9,287} & \multirow{2}{*}{49,948} & \multirow{2}{*}{9,444} & \multirow{2}{*}{40,184}\\
v0.10.4-beta & & & & & &\\[1ex]

LMAT & \multirow{2}{*}{9,767} & \multirow{2}{*}{36,339} & \multirow{2}{*}{106} & \multirow{2}{*}{36,253} & \multirow{2}{*}{2,369} & \multirow{2}{*}{12,492}\\
v1.2.1 & & & & & & \\[1ex]
MetaPhyler  & \multirow{2}{*}{0} & \multirow{2}{*}{45} & \multirow{2}{*}{0} & \multirow{2}{*}{71} & \multirow{2}{*}{0} & \multirow{2}{*}{188}\\
1.25 & & & & & & \\ [1ex]
\multirow{2}{*}{MetaCV} & \multirow{2}{*}{0} & \multirow{2}{*}{75,749} & \multirow{2}{*}{0} & \multirow{2}{*}{49,936} & \multirow{2}{*}{0} & \multirow{2}{*}{5,847}\\
& & & & & & \\[1ex]
\hline
(gdna) Wise  & \multirow{2}{*}{58,707} & \multirow{2}{*}{58,707} & \multirow{2}{*}{35,018} & \multirow{2}{*}{49,303} &  & \\
D$^{4}$RAGenS &  & & & & &\\[1ex]
\hline
\end{tabular}
\end{table}


\subsection{Bacterial Read Mapping}

The correct classification of reads to the organism, or read mapping, is a difficult task made especially challenging by the similarities of species within the same genus. Often, such species can have greater than a 95\% identity. In these instances, the majority of reads have equally strong matches to several species within the true genus, and are thus categorized by D$^{4}$RAGens as the common genus. 

The species within the genus of the spike sample A organism are distinct enough that the bulk of reads are correctly classified to the species level. Samples B and C however were created to test the precision of the algorithms and there are species within the genus with a high percentage identity. The \emph{Bacillus anthracis} present in sample C is one of three highly genetic similar pathogens in the \emph{Bacillus} genus \cite{bacillus_anthracis}.

The number of reads classified to each truth organism are shown in Table \ref{table:bacteria_read_mapping}. As predicted, algorithms perform significantly better on sample A than B or C, and are able to assign greater portions of the data to the species level. Kraken v0.10.4-beta again performs well, but the usefulness is still undermined by the extreme numbers of false positives. Again, MetaCV finds no species level matches by design, and MetaPhyler 1.25 maps under 0.5\% of the reads to the genus level. MetaScope performs exceptionally well with a read mapping rate greater than 90\%. Wise D$^{4}$RAGenS and LMAT v1.2.1 have similar read mapping performance. 

As mentioned, the D$^{4}$RAGenS methods use an optimized, smaller reference database built from RNA sequences. Since RNA is only the coding portion of the genome (i.e. the genes), and the bacterial samples contain the coding and non-coding portions, all of the unknown sequences will not be detected using the RNA database. Considering that 10-15\% of bacterial genomes are non-coding regions \cite{non_coding}  and that many reads span coding and non-coding regions, the Wise D$^{4}$RAGenS read mappings are very reasonable. 

To exhibit the read identification power of the Wise mode, the algorithm was run using the full genomic sequences (gdna) of the truth organisms as the reference database. The tallies of read mapping using the gdna are shown in the last row of Table \ref{table:bacteria_read_mapping}.  Over 93\% of sample A reads are correctly classified to the species level, and 98\% of sample B reads to the genus level, bringing the results level with MetaScope. The low false positives, reasonable run times and mapping abilities place MetaScope, D$^{4}$RAGenS, and LMAT above the other algorithms.


\subsection{Virus Identification and Read Mapping}

All seven algorithms were again run on the generated virus dataset, and results analyzed for identification and read mapping. The virus RNA dataset is about 15 times smaller than the bacterial reference.  In order to account for the decreased size, the percentage of data used in Fast mode is increased from 2 to 8\%. Despite the increase in data usage, the computation times remain fast due to the smaller number of viral organisms.

\begin{figure}[t!]
\centering
\includegraphics[width=3.7in]{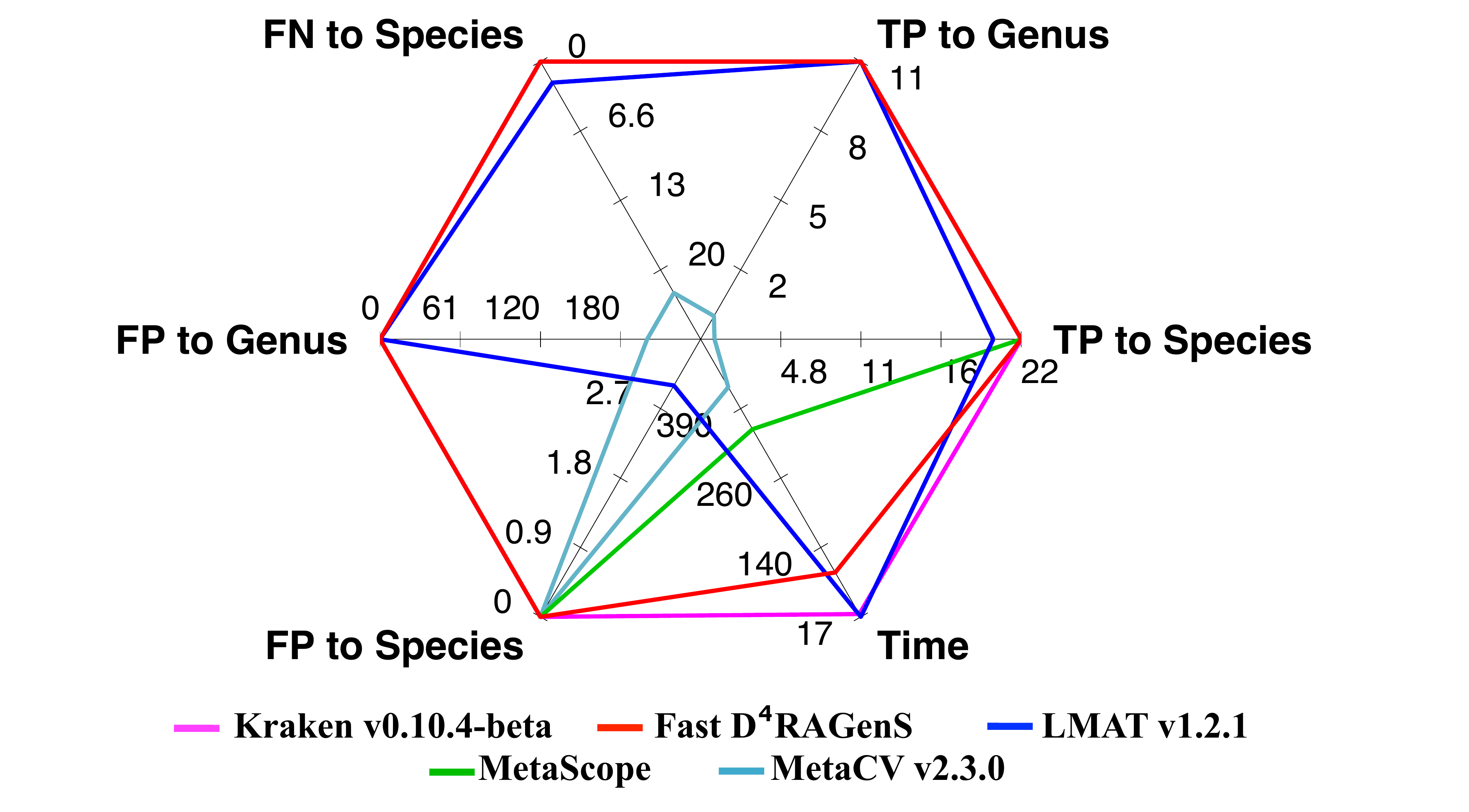}
\caption{Results of organism identification on the 22 virus \emph{in silico} Illumina test set. The  Fast D$^{4}$RAGenS algorithm scores perfect marks. }
\label{virus_id}
\end{figure}

Organism detection results displayed in Figure \ref{virus_id} show that Fast D$^{4}$RAGenS, MetaScope, and Kraken v0.10.4-beta successfully identify all 22 organisms without any false positives. LMAT v1.2.1 reported 2 false negatives and 3 false positives. MetaPhyler 1.25 and MetaPhlan v1.7.7 failed to run on the virus test set, and MetaCV v2.3.0 only identified bacterial matches. 

Read mapping results in Table \ref{table:virus_read_map} reveal the four algorithms that produced results have similar scores on this test set. Compared to bacteria, viral DNA is gene dense, with fewer non-coding regions, allowing for high read identification using only RNA.  In this case, Kraken v0.10.4-beta is believed to have no false positives due to the absence of a human background.  

\begin{table*}[!t]
\caption{Species-level read mapping results for the 22 virus \emph{in silico} dataset.}
\label{table:virus_read_map}
\centering
\begin{tabular}{l|c|cccc}
\hline
\multicolumn{1}{c|}{\textbf{Organism}}  & \textbf{Truth} & \textbf{Wise D$^{4}$RAGenS} & \textbf{MetaScope} & \textbf{Kraken v0.10.4-beta} & \textbf{LMAT v1.2.1} \\
\hline
\textbf{South American hemorrhagic fever viruses} & & & & &\\
Chapare & 961 & 614 & 947 &958 & 952\\
Guanarito & 481 & 454 & 473 & 481& 342\\
Junin & 291 & 287 & 291 &291 & 204 \\
Machupo & 196 & 190 & 195 &195 & 183\\
Sabia & 97 & 70 & 97 &97 & 97\\
\hline
\textbf{Tick-borne enchephalitis complex (flavi) viruses} & & & & &\\
Omsk hemorrhagic fever virus & 495& 327 & 495 & 494&486\\
Alkhumra hemorrhagic fever virus & 294 & 296 & 294 &294 & 0 \\
Langat Virus & 100 & 111 & 100 &100& 100\\
\hline
\textbf{Influenza} & & & & &\\
Influenza A virus & 655& 560& 605 &624& 0\\
Influenza B virus& 411&337& 381 & 401&186\\
Influenza C virus & 120&91& 115 & 118&75\\
\hline
\textbf{Filoviridae} & & & & &\\
Sudan ebolavirus & 17,277 & 16,809& 16,804 &17,274&16,976\\
Marburg marburgvirus&175&175& 175 &175&135\\
\hline
\textbf{Papillomaviridae} & & & & &\\
Human papillomavirus type 32 & 1,457&1,270& 1,457 &1,457&1,405\\
Human papillomavirus type 5 & 354&329& 350 &354&377\\
Canine papillomavirus 3 & 71&67& 71 & 71&71\\
Delta papillomavirus 1& 7 &4&7&7&7\\
\hline
\textbf{HIV and similar} & &&&&\\
Human immunodeficiency virus 1&839 &790& 737 &839&2\\
Human immunodeficiency virus 2&475&319& 473 & 475&312\\
Simian immunodeficiency virus& 88&18& 88 &88&85\\
\hline
\textbf{SARS and similar} & & & & &\\
Severe acute respiratory syndrome-related coronavirus& 273&271& 273 & 273&134\\
Human coonavirus HKU1&137&137& 137 &136&133\\
\hline
\textbf{False Negatives} & &0 & 0 & 0 &2 \\
\textbf{False Positives} & & 0 & 0 & 0 & 3\\
\hline
\end{tabular}
\end{table*}


\section{Conclusion}

Along with NGS tools, D$^{4}$RAGenS can aid in homeland security by detecting bioterror threats and disease outbreaks. In 2001 and 2011, faster analysis would have resulted in quicker implication of the source of the \emph{E. coli} and anthrax, fewer fatalities, reduced public anxiety, and less economic damage. The analysis can also be performed on an individual basis at health facilities to determine the cause of an illness before administering medications.

The techniques applied in Wise D$^{4}$RAGenS naturally lead to strain-level identification of pathogens. The pipeline can be modified to use a larger, strain specific dataset after initial identification of target organisms with Fast mode. Such an enhanced dataset would allow for detection of the minute differences between strains, leading to increased benefits for medical and threat response. Similar extensions can lead to gene detection.

Additionally, the simple string comparisons using matrix multiplications and parallel capabilities can be exploited to create an ultra-fast multiple sequence aligner. Sequence aligners show relatedness of multiple sequences and are useful in locating mutations between healthy and diseased cells (such as those associated with genetic disease and cancer). 

The basic mathematical filters used in D$^{4}$RAGenS simplify the current algorithms while maintaining a high level of accuracy. Results show the algorithm to report comparable results to  MetaScope and the publicly available LMAT v1.2.1 analysis tools. The high identification rate along with low run time and number of false positives places D$^{4}$RAGenS as a top competitor for the best genetic analysis algorithm. 


\section{Acknowledgements}
The authors are indebted to the following individuals for their technical contributions to this work: Chansup Byun, Ashley Conard, and Dylan Hutchinson.


\end{document}